\documentclass[
 reprint,
 amsmath,amssymb,
 aps,
prl,
twocolumn,
superscriptaddress,
amsmath,
amssymb,
citeautoscript,
floatfix,
]{revtex4-2}

\usepackage{graphicx}
\usepackage{adjustbox}  

\usepackage{graphicx}
\usepackage{dcolumn}
\usepackage{bm}
\usepackage{mathtools}
\usepackage[dvipsnames]{xcolor}
\usepackage{xcolor}

\usepackage[percent]{overpic}
\usepackage[letterpaper,top=2cm,bottom=2cm,left=3cm,right=3cm,marginparwidth=1.75cm]{geometry}
\usepackage{enumitem}

\begin{document}

\preprint{APS/123-QED}

\title{Contact Forces in Microgel Suspensions}
\author{Fran Ivan Vrban}
\email{fran-ivan.vrban@fmf.uni-lj.si}
\affiliation{Faculty of Mathematics and Physics, University of Ljubljana, Jadranska 19, SI-1000 Ljubljana, Slovenia}

\author{Antonio \v Siber}
\affiliation{Institut za fiziku, Bijeni\v cka cesta 46, HR-10000 Zagreb, Croatia}
\affiliation{Jožef Stefan Institute,
Jamova 39, SI-1000 Ljubljana, Slovenia}

\author{Primož Ziherl}
\affiliation{Faculty of Mathematics and Physics, University of Ljubljana, Jadranska 19, SI-1000 Ljubljana, Slovenia}
\affiliation{Jožef Stefan Institute,
Jamova 39, SI-1000 Ljubljana, Slovenia}

\date{\today}

\begin{abstract}
Within a model where micrometer-size soft colloidal particles are viewed as liquid drops, we theoretically study the contact interaction between them. We compute the exact deformation energy across a broad range of indentations and for various model parameters, and we show that it can be reproduced using either truncated superball or truncated spheropolyhedral variational shapes depending on the parameters. At large surface tensions representative of microgels, this energy is pairwise additive well beyond small indentations and can be approximated by a power-law dependence on indentation with an exponent around 2.  
\end{abstract}

\maketitle

Microgels are an important class of colloidal materials used in various applications~\cite{Karg19,Scheffold20}. In many ways, these micrometer-size soft spheres are the opposite of hard-particle sols as a key paradigm in the field: In dense suspensions, they facet and partly interpenetrate~\cite{Conley17}. Their deformed shape suggests that they may be modeled as continuous bodies, say by borrowing results from the classical elasticity of solids. Indeed, contact interaction between microgels is often described by the Hertz theory~\cite{Bergman18} which, however, only co\-vers indentations up to about 10\%~\cite{Athanasopoulou17} and does not include the surface energy.

\makeatletter
\typeout{>>> Linewidth is \the\linewidth}
\makeatother

Surface energy is an essential component of the physics of colloids and microgels are no exception. Experiments~\cite{Zhang99} and simulations~\cite{Camerin20} show that microgels stabilize the oil-water interface, imply\-ing that they have an effective surface tension $\gamma_F$ even though their surface is not as sharp as in mole\-cu\-lar liquids and solids. This has important implications. The surface ener\-gy of a sphere of radius $R_*$ scales as $\gamma_F R_*^2$, whereas its bulk ener\-gy is proportional to $KR_*^3$ where $K$ is the bulk modulus. The surface-to-bulk energy ratio reads 
\begin{align}
    \frac{\gamma_F R_*^2}{KR_*^3}=\frac{\ell_S}{R_*},
    \label{eq:ellsR}
\end{align}
where $\ell_S=\gamma_F/K$ is the elastocapillary length~\cite{Bico18}. In microgels, $\ell_S$ is between 10 and 100~$\mu$m~\cite{comment1} whereas $R_*$ is usually a few 100~nm. Thus $\ell_S/R_*>1$, which shows that the mechanics of micro\-gels is governed by surface tension. In addition, the small size of these particles means that the stress within them is uniform rather than distributed. These two conclusions are incompatible with the assumptions of the Hertz theory, which builds on classical elasticity alone, but are included in the li\-quid drop model~\cite{Riest15,Doukas18}. 
In this model, deformable colloidal particles are viewed as compressible drops with a differential affinity to each other and to the solvent, allowing one to distinguish between the poor- and good-solvent regimes. In the special case of incompressible particles lacking
differential affinity, the liquid drop model reduces to
the foam theory of emulsions~\cite{Lacasse96a} known, e.g., for the Morse--Witten drop-drop interaction~\cite{Morse93}.

So far, the liquid drop model was primarily explored numerically~\cite{Riest15,Doukas18}, its high-density variant explaining the existence of exotic crystals formed by nanocolloidal micelles~\cite{Ziherl00}. Here we use it to theoretically study the contact interaction between microgels and other soft $\mu$m-size particles. We show that the exact deformation energy can be reproduced by two variational shapes and that in the large-tension regime representative of microgels, it reduces to a simple power law and is pairwise additive. 

\paragraph{The liquid drop model.} We start with the phenomenolo\-gical free energy of a drop:
\begin{align}
F =\frac{V_0}{\chi_T}\left(\frac{V}{V_0}-1-\ln\frac{V}{V_0}\right)+\gamma_FA_F+\frac{\gamma_CA_C}{2}.
\label{eq:LDM1}
\end{align}
The bulk term represents the Murnaghan equation of state with a pressure that vanishes when the drop volume $V$ equals the reference volu\-me $V_0$; $\chi_T$ is the compressibility. In the surface terms, $A_F$ and $A_C$ are the areas of the non-contact domain and the contact zones, respectively, and $\gamma_F$ and $\gamma_C$ are the corresponding tensions. The factor of 1/2 in the last term appears because each contact zone is shared by two drops. 

\begin{figure}[t]
  \centering
  \vspace*{-2.5mm}
  \includegraphics[width=1\linewidth]{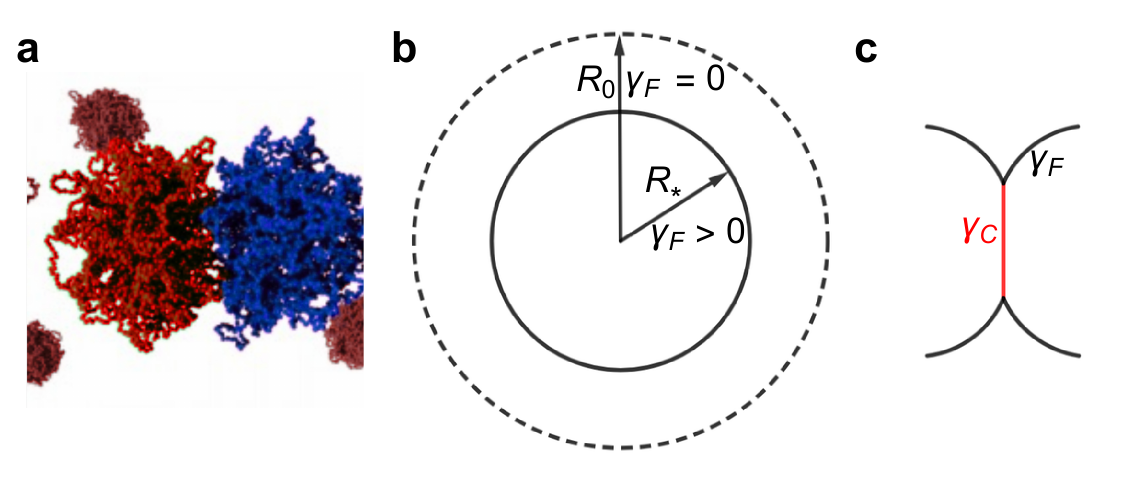}
  \vspace*{-5mm}
  \caption{(Color online) 
    Snapshot of two interacting microgels obtained using monomer-resolved numerical simulations (a; image courtesy of L. Rovigatti and E.~Zaccarelli). Panel b shows the relationship between the reference and the resting drop, and panel c emphasizes the difference between contact and non-contact tension.
  }
  \label{fig:sketch}
\end{figure}

The first dimensionless parameter of the model is the reduced Egelstaff--Widom length~\cite{Egelstaff70,Riest15} 
\begin{align}
    \Psi=\frac{2\gamma_F\chi_T}{R_0},
\end{align}
which encodes the relative vo\-lu\-me decrease due to the Laplace pressure $2\gamma_F/R_0$, whereby the re\-fe\-rence radius $R_0=(3V_0/4\pi)^{1/3}$ shrinks to the resting radius $R_*$ (Fig.~\ref{fig:sketch}; Supplemental Material~\cite{Supplementary}, Sec.~I); note that $\Psi=2\ell_S/R_* \times R_*/R_0$. The second parameter is the tension ratio
\begin{align}
    \omega=\frac{\gamma_C}{2\gamma_F}.
\end{align}

\paragraph{Deformation free energy.} We first compute the equi\-librium drop shapes numerically using Surface Evolver~\cite{Brakke92} as elaborated in End Matter and Supplemental Material~\cite{Supplementary}, Sec.~III; given that our model is a coarse-grained description of soft colloidal particles, we disregard a possible fine curvature of the contact zones~\cite{Thomson87} and assume that they are flat. 
Thus we obtain the drop deformation free energy $\Delta F$ defined rela\-tive to the resting drop. This is done for a wide range of reduced Egelstaff--Widom lengths $\Psi$; $\Psi\ll1$ and $\Psi\gg1$ are referred to as the small- and the large-tension regime, respectively, the latter corresponding to microgels. We exa\-mi\-ne tension ratios $\omega$ both lar\-ger and smaller than unity, initially for drops in the simple cubic (SC) lattice where the coordination number $z=6$.

\begin{figure}[t]
    \hspace*{-0.0cm}
    \includegraphics[width=1.005\linewidth]{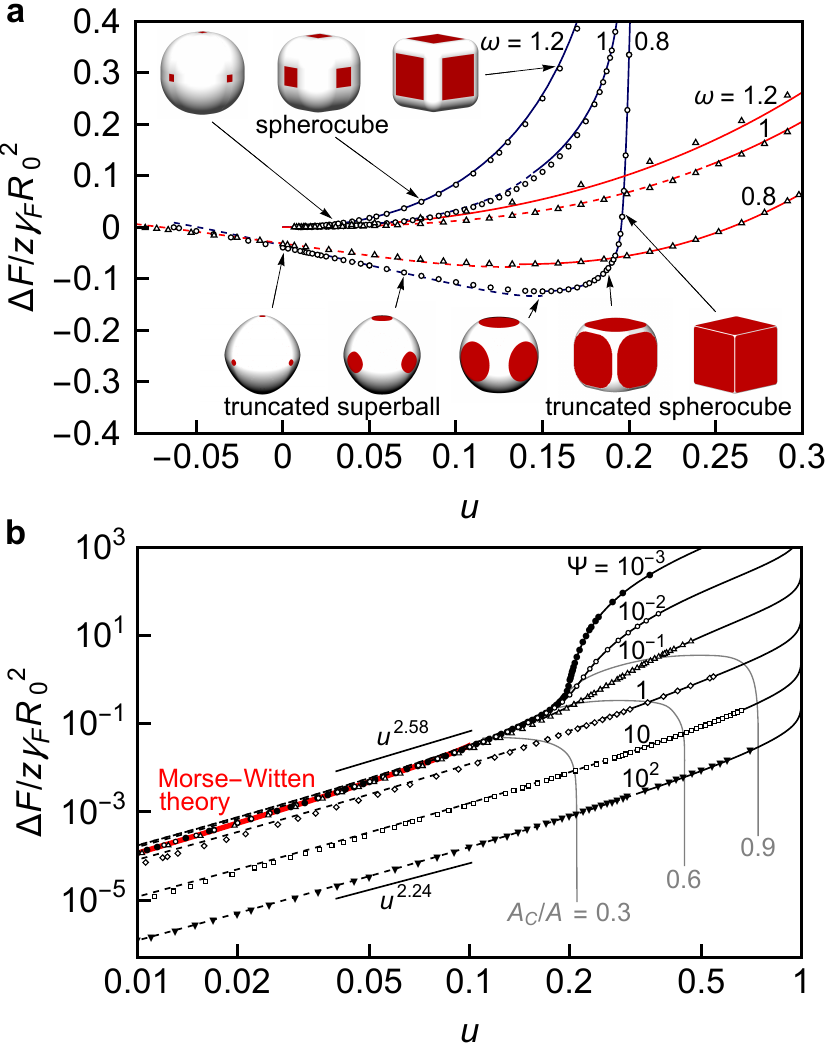}
    \caption{(Color online) 
Deformation free energy per contact $\Delta F/z\gamma_FR_0^2$ in $\Psi=10^{-3}$ (circles; blue [dark gray]) and $\Psi=1$ (triangles; red [light gray]) drops in SC lattice vs.~dimensionless engineering indentation $u$~(a), comparing exact (symbols) and variational results (solid lines) at $\omega=0.8,1,$ and 1.2. Insets show a few representative shapes at $\Psi=10^{-3}$ and $\omega=0.8$ and 1.2. In panel b, we plot exact (symbols) and variational (truncated superball: dashed lines, truncated spherocube: solid lines) $\Delta F/z\gamma_FR_0^2$ vs.~$u$ for $\Psi=10^{-3}, 10^{-2},\ldots, 10^2$ at $\omega=1$. Also included are isolines showing the ratio of contact and total area $A_C/A=0.3, 0.6,$ and $0.9$ (light gray lines) and the Morse--Witten result (thick red [light gray] line). 
    }
    \label{fig:DeltaF}
\end{figure}

Figure~\ref{fig:DeltaF}a shows $\Delta F$ as a function of dimensionless engineering indentation 
\begin{align}
    u=\frac{h}{R_*},
\end{align} 
where $h$ is the displacement of the contact zone toward drop center. We present two datasets, one for $\Psi=10^{-3}$ and the other for $\Psi=1$, and we compare attractive drops with $\omega=0.8$ to neutral and repulsive ones at $\omega=1$ and 1.2, respectively.

The two sets of data in Fig.~\ref{fig:DeltaF}a are qualitatively similar. At $\omega=0.8$, the drop-drop inter\-action is attractive at small $u$; the attractive regime extends to negative $u$ where the drops are pulled apart but form cohesive capillary bridges with each other. At $u$ beyond the mini\-mum lies the repulsive regime where the drops are increasingly more faceted as illustrated by the $\Psi=10^{-3}$, $\omega=0.8$ sequence of variational shapes included in Fig.~\ref{fig:DeltaF}a and described below. This sequence co\-vers both partial faceting where the contact zones are small as well as complete faceting where the drop is essentially polyhedral and the ratio of 
contact and total area $A_C/A$ shown by isolines in Fig.~\ref{fig:DeltaF}b approaches unity. In the $\omega=1$ and 1.2 drops, the deformation free energy is repulsive at all $u$, its magnitude at $\omega =1.2$ being larger than at $\omega=1$ as expected; here too we include a few best-fit {(i.e., lowest energy) variational shapes shown in the $\Psi=10^{-3}$, $\omega=1.2$ insets to Fig.~\ref{fig:DeltaF}b.  

\paragraph{Variational shapes.} The form of drops in contact seems deceptively simple (Supplemental Material \cite{Supplementary}, Sec.~III), and our next goal is to better understand it using variational approxi\-mations. 
The two ans\" atze explored here are truncated superballs and sphero\-polyhedra (Supplemental Material \cite{Supplementary}, Secs.~V and VI; 
insets to Fig.~\ref{fig:DeltaF}a). Our choice of variational shapes is based on four guidelines: They should be (i)~derived from the sphere so as to capture the small-indentation regime, (ii)~compatible with an arbitrary arrangement of neighbors, at least at small indentations, and (iii)~specified by only a few parameters. Finally, they should (iv)~satisfy as best as possible the force-balance condition at the edge of the contact zone, which gives $\cos\theta = \omega$ where $\theta$ is the contact angle so that in attractive ($\omega<1$) drops $\theta<\pi$ whereas in repulsive ($\omega>1$) ones $\theta=\pi$. 

Unlike the Z-cone model~\cite{Hutzler14}, our shapes are continuous and closed and they include the truncated sphere (Supplemental Material \cite{Supplementary}, Sec.~IV) as a special case. The 6-pole super\-ball compatible with the SC lattice 
is defined by 
\begin{align}
    |x|^m+|y|^m+|z|^m=R^m.
    \label{eq:superball}
\end{align}
For $m<2$, the poles are pointed whereas for $m>2$ they are squashed. The superball is truncated by removing pole caps and the thus formed contact zones are more or less circular as long as $m\approx2$; the contact angle $\theta<\pi$. Equation~(\ref{eq:superball}) can be modified to construct superballs with a different number and arrangement of poles~\cite{Onaka06}. 
The second shape studied is the spheropolyhedron, i.e., the polyhedron with rounded edges and vertices with the same radius of curvature~$R_{\rm edge}$. The shape of each contact zone is the same as that of the parent face and the contact angle $\theta=\pi$. In the truncated variant where plane-parallel slices are removed from the faces, $\theta<\pi$ and the contact zones are polygons with rounded vertices. 

The relevance of each variational model is determined by how well its deformation ener\-gy agrees with the exact one (Supplemental Material~\cite{Supplementary}, Sec.~VIIA). As evidenced by Fig.~\ref{fig:DeltaF}a, at $\omega\lesssim1$ the exact $\Delta F$ is reproduced very well by a combination of truncated superball and truncated spherocube at $u<u_0$ and $u>u_0$, respectively, with $u_0\approx0.15$ at $\Psi=10^{-3}$ and $u_0\approx0.25$ at $\Psi=1$. On the other hand, at $\omega\gtrsim1$ the best-fit shape is the spherocube at all $u$. Equally good is the agreement of the variational shapes with the exact $\Delta F$ at $\omega=1$ and $\Psi=10^{-3},10^{-2},\ldots 10^2$ shown in Fig.~\ref{fig:DeltaF}b, which spans many orders of magnitude. A more detailed insight into the quality of the fits is provided by Fig.~\ref{fig:discrepancy} showing the rela\-tive diffe\-rence of the two ener\-gies. At small $\Psi$, the deviation of the variational $\Delta F$ reaches about 30~\% at very small indentations but falls off rapidly and vani\-shes in the complete faceting regime, whereas at $\Psi=10^2$ it peaks at about 10~\% at $u\approx0.03$. 
\begin{figure}[ht]
\includegraphics[width=1\linewidth]{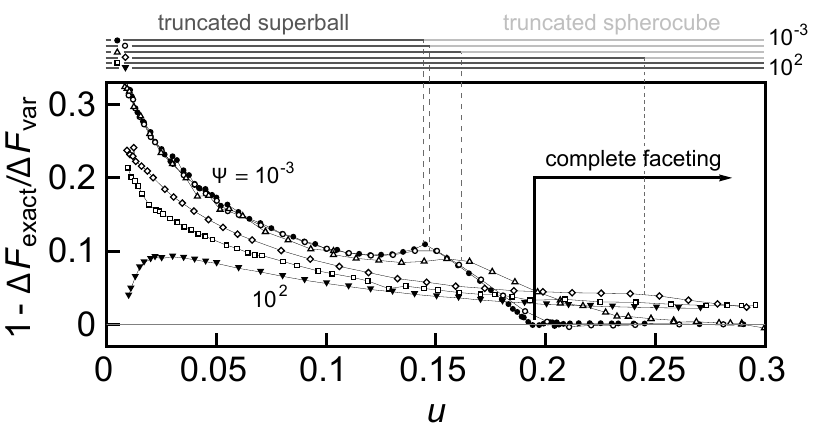}
    \caption{
  Relative difference of variational and exact deformation free ener\-gies for $\Psi=10^{-3},10^{-2},\ldots 10^2$ (top to bottom in the left half of the plot) and $\omega=1$; points are connected for clarity. The arrow shows the onset of complete faceting for $\Psi=10^{-3}$ and $10^{-2}.$ The dark and light gray bars above the plot indicate the ranges of $u$ for which the best-fit variational shape is the truncated superball and the truncated spherocube, respectively, at a given $\Psi$. For $\Psi=10$ and $100$, the transition between the two shapes is at $u>0.3$.}
    \label{fig:discrepancy}
\end{figure}
Note that our variational shapes are described by 
just 2 parameters: In the superball, we vary the exponent $m$, the ratio $R/R_*$, and the degree of truncation, whereas spheropolyhedra are controlled by the dimensionless radius of the rounded parts $R_{\rm edge}/R_*$ and the degree of truncation.
 
\paragraph{Scaling laws.} Figure~\ref{fig:DeltaF}b emphasizes many features of the deformation free energy across a wide 
range of reduced Egelstaff--Widom lengths and indentations. Specifically, it 
shows that (i)~for $\Psi\lesssim10^{-1}$, the small-inden\-tation $\Delta F$ is independent of $\Psi$ and perfectly consistent with the Morse--Witten theory~\cite{Morse93} whereas (ii)~at the onset of complete faceting at $u\gtrsim0.2$, $\Delta F$ increases in step-like fashion; (iii)~at $u\gtrsim0.3$, $\Delta F\propto1/\Psi$ for all $\Psi$; and (iv)~at $\Psi\gtrsim1$, $\Delta F$ is proportional to $1/\Psi$ for all $u$. These scaling laws can be obtained analytically using suitable simplifications (Supplemental Material \cite{Supplementary}, Sec.~VIII). Finally, we note that at $\Psi\gtrsim1$ and $\omega\geq1$
\begin{align}
\frac{\Delta F}{z\gamma_F R_0^2}\approx\frac{3u^{\alpha}}{\Psi},
\label{eq:pwr}
\end{align}
where $\alpha$ weakly depends on $\Psi$; at $\Psi=10$ and $\omega=1$, e.g., $\alpha=2.28$ (Supplemental Material~\cite{Supplementary}, Sec.~IX). Equation~(\ref{eq:pwr}) holds to a good approximation for all $u\lesssim0.4$, and is of practical importance because it describes the drop-drop interaction in the regime pertaining to microgels.

\paragraph{Many-body effects.} The Morse--Witten theory 
shows that at small tensions, the drop-drop interaction is markedly many-body even at small indentations. To see whether this also holds at large tensions, we compare $\Delta F/z\gamma_FR_0^2$ in the $z=6$ SC lattice and in the $z=4$ diamond-cubic (DC) lattice; the discussion is limi\-ted to $u\lesssim0.36$ where a drop in the DC lattice presses on the 4 nearest neighbors but not on the 12 next-nearest ones. The many-body effects are quantified using the rela\-tive difference of the deformation free ener\-gies per contact defined by
\begin{align}
    \Xi=1-\frac{(\Delta F/z)_{z=4}}{(\Delta F/z)_{z=6}}.
\end{align}
If the interaction is pairwise additive, $\Xi=0$. 

Figure~\ref{fig:ManyBodyX} shows $\Xi$ obtained from the exact deformation free energies at $\omega =1$ and $\Psi=10^{-3},10^{-2},\ldots10^2$. At $\Psi=10^{-3}$, $\Xi$ is positive and increases with $u$, nicely agreeing with the Morse--Witten result at $u\lesssim0.1$. As $\Psi$ is increased to $10^{-1}$, $\Xi$ remains qualitatively similar but decreases in magnitude, which indicates weaker many-body effects. In large-tension drops with $\Psi=10$, $\Xi$ falls to less than $0.03$ at all indentations considered, implying that the contact interaction is essentially pairwise additive; this holds at all $\Psi\gtrsim10$. The values of $\Xi$ computed using the truncated-spheropolyhedron model are systematically smaller than the exact ones but their overall trend is the same. The deviation is largest at small $u$, where $\Xi$ involves a difference of two small numbers and the variational models are less accurate than at large $u$ as seen in Fig.~\ref{fig:discrepancy} (Supplemental Material~\cite{Supplementary}, Sec.~VIIB).
\begin{figure}[t]
    \includegraphics[width=1\linewidth]{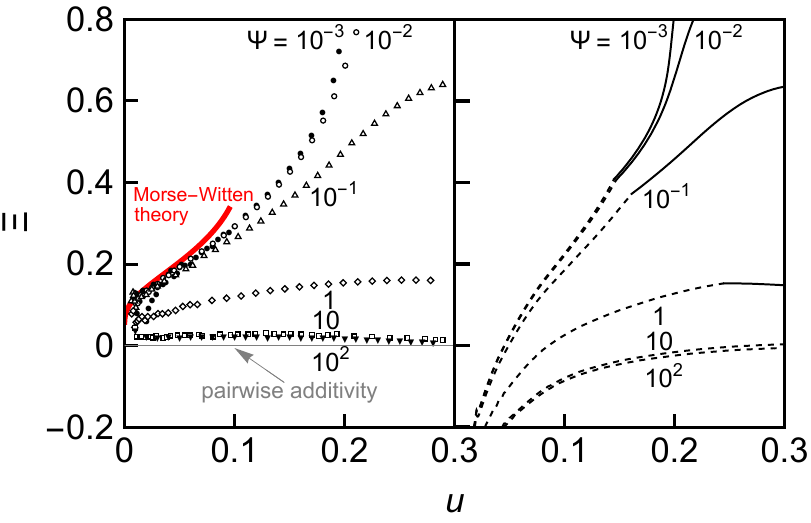}
    \caption{(Color online) 
  Relative difference of the deformation free ener\-gies in the DC and SC lattice $\Xi$ at $\omega=1$ for $\Psi=10^{-3},10^{-2},\ldots10^2$. In the left panel, symbols represent exact results and the thick red [light gray] line is the prediction of the Morse--Witten theory. The right panel shows variational results: Dashed and solid lines indicate the domains in which the best-fit models for the SC lattice are the cubic superball and the spherocube, respectively, whereas the tetrahedral superball provides the best fit for the DC lattice for all $u\in[0,0.3]$.}
    \label{fig:ManyBodyX}
\end{figure}

The somewhat unexpected pairwise additivity of the contact interaction at large $\Psi$ appears to originate in the bulk term dominated by work done against the refe\-rence pressure, the relative pressure increase due to indentation being small, and in the drop radius virtually unaffected by indentation (Supplemental Material~\cite{Supplementary}, Sec.~IVD). These related effects arise 
because the resting state is determined by the balance of strong contractile surface-tension forces and a large drop pressure, 
the latter being independent of indentation to leading order. In contrast, in the small-tension regime the resting state is defined prima\-ri\-ly by the volume constraint, and thus any drop-drop contact necessarily leads to a global deformation so that the interaction is many-body. The above insight into the pairwise-additive nature of drop-drop interaction at large $\Psi$ should be further explored, say using the criterion based on the Cauchy relation~\cite{Reinke07}. Such a test would identify the conditions where pairwise additivity is applicable universally, i.e., irrespective of the symmetry 
of the local geometry.

\paragraph{Discussion.} 
This work offers a new view of contact forces between large-tension drops representative of microgels. We show that within the liquid drop model, the deformation free ener\-gy is given by an effective power law $\Delta F\propto u^\alpha$ with $\alpha\approx2$ (Supplemental Material~\cite{Supplementary}, Sec.~IX), which holds well beyond small indentations---and that it is pairwise additive. These two features of the drop-drop interaction are also important because they {\sl a posteriori} justify the assumptions of theoretical studies that postulate pairwise additivity and involve similar interparticle potentials (e.g., Refs.~\cite{Pamies09,Zhang09,Bergman18}). 
Our findings uphold the relevance of these theories for suspensions of deformable $\mu$m- and nm-size particles.

Our second result is in showing that the truncated superball and spheropolyhedron shapes nicely capture the main morphological features of drops in the repulsive and the attractive regime, respectively, cove\-ring all stages of faceting. These variational shapes offer an evi\-dent computational advantage at a reasonable cost in accuracy. Since they are equally applicable to ordered and disordered local geometries, these shapes can be employed for efficient computational studies of large assemblies of microgels where each particle would be represented by a spheropolyhedron based on its Voronoi cell; the geometry of such a scheme can be visualized by the network of Plateau borders in wet foam~\cite{Kraynik06}. From this perspective, our work paves the way to theoretical investigations of a broad class of materials including emulsions and suspensions of soft particles such as microgels, polymeric droplets, and thin-shell microcapsules at scales permitting studies of collective and emergent phenomena, possibly using more refined versions of the model with, e.g., curvature-dependent surface non-contact tension~\cite{Tolman49} and alternative types of bulk and surface free energy.

\acknowledgments

We thank C.~N.~Likos, 
M.~Kandu\v c, F.~Scheffold, and D.~Vlassopoulos for helpful discussions, and L. Rovigatti and E.~Zaccarelli for the image in Fig.~\ref{fig:sketch}a. F.~I.~V.~and P.~Z.~acknowledge support from EU MSCA Doctoral Network QLUSTER, Grant Agreement 101072964. A.~Š.~acknowledges the support of the project EMONA-P financed by the European Union through Croatia's National Recovery and Resilience Plan 2021-2026. P.~Z.~acknow\-ledges financial support from the Slovenian Research Agency (research core funding No.~P1-0055). 

\nocite{SierraMartin11,LietorSantos11,Li14}

\bibliographystyle{apsrev4-1}
\bibliography{LDM}

\section{End Matter}

{\bf Exact drop shapes.} The exact minimal-energy drop shapes were obtained numerically using the Surface Evolver package~\cite{Brakke92}. We employed a scheme where drop-drop contacts lie on planes bisecting center-to-center distances as if the drop were confined to a polyhedral Wigner--Seitz cell. In this scheme, the contact zones are flat and their fine curvature known from, e.g., the Kelvin foam~\cite{Thomson87} is disregarded; given that our drops are used as models of microgels and that the notion of the contact zone as a 2D object itself is an idealization, this approximation is justified.  

At a given $\Psi$ and $\omega$, we computed the deformation free energy across a range of indentations $u$ as follows:
\begin{enumerate}[label=(\roman*)]

\item First we determined the resting shape of the drop in absence of any constraints. This was done by starting from a cubic initial guess, which was then refined and retriangulated using equiangulation and vertex ave\-ra\-ging routines several times so as to ensure that the precision of the obtained resting drop energy is high enough. This requirement is important because the drop deformation energy is defined relative to the resting state. Also computed in this step was the reference radius $R_*$. 

\item In the second step, the drop was confined to a polyhedron corresponding to the SC or DC lattice at a certain engineering indentation $u$. The facets that fell beyond a constraint were fixed to the constraint and assigned the value of the contact surface tension, and then the drop shape was recalculated. After that, the facets lying on constraints were detached from them and the mesh was relaxed; this was followed by reattachment of the facets at constraints to them. This dynamical allocation of facets to constraints, performed together with the remeshing involving refinement, vertex averaging, and removal of small triangles, was repeated until the energy converged. A ty\-pi\-cal final mesh had about 15000 vertices, 30000 facets and 45000 edges. 

\item After the equilibrium shape and energy at a given indentation $u$ were obtained, indentation was increased and step~(ii) was repeated.

\end{enumerate}

When convergence was slow or the shape lingered far from the minimum, additional means were applied as needed so as to promote convergence. In case of attractive interactions ($\omega<1$), we started with the minimal-energy shape at $\omega=1$ and then $\omega$ was gradually decreased to the desired value. In severely constrained drops which almost completely fill out the confining polyhedron, additional mesh remodeling steps were introduced so as to ensure that the neighboring contact zones remained separated by a non-contact strip.

{\bf Variational shapes.} Our variational shapes are controlled by 2 parameters (exponent $m$ and ratio $R/R_*$ in superballs where the latter also defines the degree of truncation; $R_{\rm edge}/R_*$ and the degree of truncation $\eta$ in spheropolyhedra). In both shapes, the deformation free energy is an explicit but complicated and nonlinear function of the variational parameters which cannot be mini\-mized analytically. At a given indentation $u$, the minimal-energy values of the variational para\-meters subject to constraints (Supplemental Material~\cite{Supplementary}, Sec.~VI) were computed using a nume\-rical solver in Wolfram 14.3. The solver searched the allowed parameter space by employing opti\-mization routines that enforce the constraints so as to avoid local minima and to prevent non-physical variational shapes. This procedure was repeated as the dimensionless indentation $u$ was increased in small increments, and the energy-minimizing values of the variational parameters were tracked in order to ensure continuous shape deformation as illustrated, e.g., in Fig.~S15 of the Supplemental Material~\cite{Supplementary} which shows the dimensionless ratio $R_{\rm edge}/R_*$ and the degree of truncation in spherocube vs.~$u$ at various $\Psi$ and $\omega=1$.

\end{document}